# Accelerated Algorithms for Source Orientation Detection and Spatiotemporal LCMV Beamforming in EEG Source Localization

Ava Yektaeian Vaziri, Bahador Makkiabadi
Department of Biomedical Engineering, Tehran University of Medical Sciences
Tehran, Iran
yektaeianvaziri@gmail.com

*Abstract*— **This paper illustrates the development of two efficient source localization algorithms for electroencephalography (EEG) data, aimed at enhancing real-time brain signal reconstruction while addressing the computational challenges of traditional methods. Accurate EEG source localization is crucial for applications in cognitive neuroscience, neurorehabilitation, and brain-computer interfaces (BCIs). To make significant progress toward precise source orientation detection and improved signal reconstruction, we introduce the Accelerated Linear Constrained Minimum Variance (ALCMV) beamforming toolbox and the Accelerated Brain Source Orientation Detection (AORI) toolbox. The ALCMV algorithm speeds up EEG source reconstruction by utilizing recursive covariance matrix calculations, while AORI simplifies source orientation detection from three dimensions to one, reducing computational load by 66% compared to conventional methods. Using both simulated and real EEG data, we demonstrate that these algorithms maintain high accuracy, with orientation errors below 0.2% and signal reconstruction accuracy within 2%. These findings suggest that the proposed toolboxes represent a substantial advancement in the efficiency and speed of EEG source localization, making them well-suited for real-time neurotechnological applications.**

***Keywords—EEG source localization[1], Beamforming, neural signal processing, LCMV, Accelerated algorithms, Source orientation detection, Recursive calculations, Brain-computer interface (BCI)***

## I. INTRODUCTION

Accurate source localization in electroencephalography (EEG) is fundamental for understanding brain function and neural dynamics, making it essential in cognitive neuroscience, neurorehabilitation, and brain-computer interface (BCI) applications. Despite the critical role of EEG in these fields, high speed processing remains a significant challenge due to the high computational demands of current source localization algorithms. As applications continue to demand both speed and precision, the development of accelerated algorithms for source orientation detection and beamforming has become increasingly important.

Traditionally, Linearly Constrained Minimum Variance (LCMV) beamforming has been the go-to technique for EEG source localization. This approach is favored for its ability to spatially filter signals, enhancing the signal-to-noise ratio by suppressing noise and interference. However, LCMV beamforming methods can be computationally intensive, especially when applied to high-density EEG systems with numerous channels [1]. The computational complexity of covariance matrix calculations and matrix inversions, coupled with the need for accurate estimation of brain source orientation, makes fast processing difficult.

To address these challenges, we propose a set of accelerated algorithms aimed at improving the efficiency of both LCMV beamforming and source orientation detection. Specifically, we have developed two MATLAB toolboxes [2]: ALCMV and AORI. The ALCMV toolbox expedites the LCMV beamforming by initiating recursive calculations as soon as we have acquired EEG samples equivalent to the number of electrode channels—necessary for achieving a full-rank matrix—we can significantly enhance processing speed. This approach enables faster source localization while maintaining accuracy [3].

In the LCMV algorithm, electrical signals generated by brain sources in specific directions propagate through various layers of the brain and ultimately reach the EEG recording electrodes. If the source activity is denoted by S(t) and the estimated source activity by $\hat{S}(t)$, we employ beamforming techniques to isolate and reconstruct the source signals from the noisy background [4]. The LCMV beamformer is formulated as a spatial adaptive filter that minimizes the output variance, enhancing its ability to localize sources. Mathematically, this is expressed in Equation 1.

$$w(r) = R^{-1}L(r)[L^T(r)R^{-1}L(r)]^{-1}$$

*Equation 1*

where w(r) is the beamforming weight vector, R is the covariance matrix of the transposed EEG data, and $L$ is the lead field matrix, typically obtained through MRI imaging and forward problem solutions [5]. The source estimate S^(t) is given by Equation 2.

$$\hat{S}(r,t) = w^T(r)y(t)$$

1 electroencephalogram

*Equation 2*

where y(t) is the recorded EEG signal [6]. $w^T(r)$ is the transpose of w(r) from Equation 1.

### A. Covariance Calculation and Efficiency

The covariance matrix is a key computational component when working with EEG data, particularly in the context of LCMV beamforming. For k channels and N recorded samples, the EEG data matrix typically has dimensions k×N. However, the covariance matrix derived from the transposed EEG data is a much smaller k×k matrix, reducing the computational burden and providing substantial efficiency. The covariance between EEG channels j and k can be computed in Equation 3 [7].

$$a_{jk} = \frac{1}{N-1}\sum_{i=1}^{N}(x_{ij} - \overline{x_j}) \times (x_{ik} - \overline{x_k})$$

*Equation 3*

where $\overline{x_j}$ and $\overline{x_k}$ are the means of the $j$-th row and k-th column, respectively. This computational efficiency is one of the primary advantages of the LCMV beamformer, allowing for precise source localization without excessive computational load.

### B. Source Orientation Estimation

Estimating the orientation of brain sources often involves utilizing 3D MRI images of the individual. However, this process can be computationally demanding. To improve efficiency, we developed the AORI toolbox, which reduces the dimensionality of the orientation problem from three to one by leveraging the lead field matrix and the inverse covariance matrix. The optimal direction $\eta_{opt}(r)$ is calculated by maximizing the output power. Mathematically, the optimal orientation is given in Equation 4.

$$\eta_{opt}(r) = \vartheta_{min}\{L^T(r)R^{-1}L(r)\}$$

*Equation 4*

where $\eta_{opt}(r)$ represents the eigenvector corresponding to the smallest eigenvalue. The reduction in dimensionality can lead to faster calculations and more efficient source localization.

### C. Literature Review

Existing EEG source localization methods, such as minimum norm estimation and traditional LCMV beamforming, have demonstrated their effectiveness in various applications. However, these methods often struggle with fast processing due to their reliance on extensive matrix computations. For example, minimum norm estimation techniques tend to require significant computational resources, making them less suitable for fast applications [8]. Similarly, conventional LCMV beamforming can become computationally prohibitive as the number of EEG channels increases, leading to longer processing times and potential delays in source localization [9]. Our approach addresses these limitations by introducing novel algorithms that reduce the computational load without compromising accuracy. The recursive calculation method implemented in the ALCMV toolbox and the dimensionality reduction provided by the AORI toolbox offer significant improvements over traditional methods [10]. By building on the strengths of LCMV beamforming while mitigating its weaknesses, our work represents a meaningful advancement in the field of EEG source localization [4].

### D. Significance and Innovation

The primary innovation of this study lies in the combination of recursive calculations and dimensionality reduction to accelerate source localization in EEG. Unlike previous approaches that treat the problem in a static manner, our methods dynamically adapt to the incoming EEG data, enabling faster calculations. The ALCMV toolbox optimizes the LCMV beamforming process, reducing the time required to achieve full-rank matrix conditions. At the same time, the AORI toolbox transforms the complex three-dimensional orientation problem into a more manageable one-dimensional problem, further enhancing computational efficiency. These innovations are not only theoretical but have been implemented in MATLAB toolboxes designed for practical use in real-time applications. By validating our algorithms with simulated EEG signals using the EEGg toolbox [11], we have demonstrated that these methods can achieve substantial gains in processing speed while maintaining the precision necessary for accurate source localization.

### E. Applications and Broader Impact

The accelerated algorithms developed in this study hold significant potential for a wide range of EEG applications. In neurorehabilitation, for instance, fast and accurate source localization is crucial for real-time feedback in brain-computer interface systems. Similarly, in clinical diagnostics, such as seizure detection and monitoring, the ability to quickly identify the origin of abnormal brain activity can be life-saving. Our toolboxes provide the better spatial resolution without sacrificing accuracy, making them valuable assets for both research and clinical practice.

### F. Objective and Structure

This study aims to develop and evaluate accelerated algorithms for source orientation detection and LCMV beamforming in EEG source localization. The rest of the paper is structured as follows: section 3 describes the methodology behind our accelerated algorithms, detailing the recursive calculations and dimensionality reduction techniques. Section 4 presents the results of our simulations and performance evaluations, while Section 5 discusses the implications of our findings and potential future directions.

## II. METHOD

### A. Covariance calculation in ALCMV beamforming

To address inverse covariance requirement reliably, we adopted the Miller computational method for recursively calculating the inverse covariance. In the ALCMV method, covariance was initially calculated using the addition and subtraction technique. Subsequently, inversion was performed using the general inverse method, implemented through Miller's approach. In Figure 1, covariance using the addition and subtraction method contains first samples (ns) that are equal to the number of channels (k) and calculate the covariance using the standard method, which gives us the initial covariance ($C_1$). After recording another cy (cycle) samples now we have ns+cy samples, we separate the first cy samples of the EEG data and the last cy samples of the EEG data and name them ($D_1$) and ($\acute{D}_1$), respectively. Then, we calculate the covariance of the transposed $D_1$ and $\acute{D}_1$ and name them $R_1$ and $\acute{R}_1$. Finally, we use Equation 4 to calculate the second updated covariance $C_2$. The Equation 4 can be extended in Equation 5 [7].

$$C_2 = C_1 - R_1 + \acute{R}_1$$

*Equation 5*

$$C_{n+1} = C_n - R_n + \acute{R}_n$$

*Equation 6*

It is important to note that the number of columns used in the initial calculations (ns) must ensure that the matrix is full rank. In other words, ns should be at least equal to the number of electrode signal channels (k). For the remaining data, however, the covariance can be updated incrementally (cy=1), vector by vector or any value less than (ns/2).

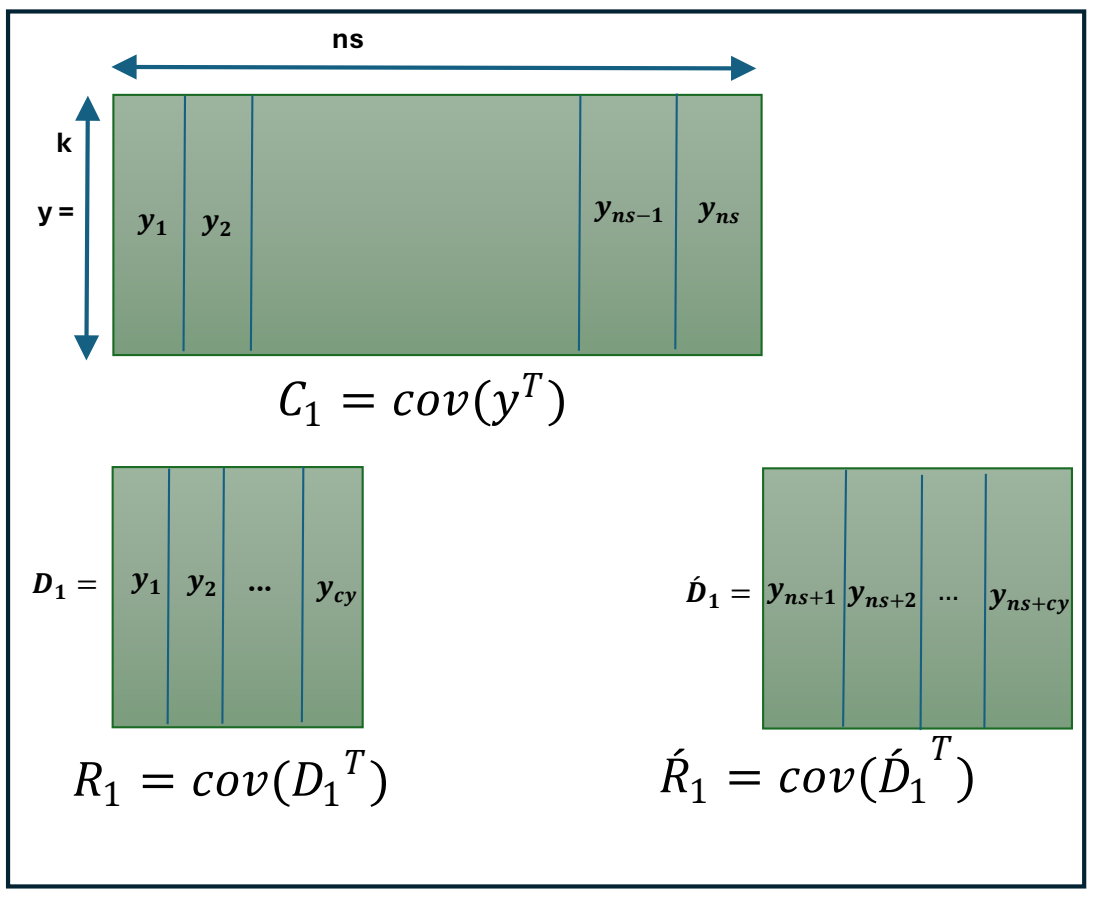

*Figure 1 Covariance calculation in ALCMV beamforming*

## B. Inverse covariance calculation with General Miller (GM) in ALCMV Beamforming

To calculate the inverse covariance matrix efficiently, we employed the General Miller (GM) method, which is known for its robustness in handling high-dimensional data. The GM approach in Equation 7 leverages iterative refinement to achieve accurate estimates, making it particularly suitable for cases where standard inversion techniques may fail due to numerical instability [12].

$$C_{k+1}^{-1} = C_k^{-1} - v_k C_k^{-1} E_k C_k^{-1}$$

*Equation 7*

### B.1 Miller Algorithm

Each time covariance is calculated using the addition and subtraction method, before the final addition and subtraction, in Miller method the inversion is performed specifically for the addition and subtraction. By combining the two methods, we proceed step by step for recursive and accelerated calculations.

### B.2 Miller's equation

To apply the Miller method to Equation 6, we first denote the current covariance ($C_n$) matrix as $G$ (Equation 8). Then, we define $-R_n + \acute{R}_n$ as $H$ (Equation 9). Thus, the updated covariance matrix ($C_{n+1}$) is expressed as $G + H$ (Equation 10).

$$C_n = G$$

*Equation 8 step one in Miller equation*

$$H = -R_n + \acute{R}_n$$

*Equation 9 step two in Miller equation*

$$C_{n+1} = G + H$$

*Equation 10 step three in Miller equation*

## C. General Miller equation

In the general Miller method, the matrix $H$ is transformed into a sum of rank-one matrices. This involves decomposing $H$ into a summation of rank-one components, where each component is a product of single column vectors and their corresponding row vectors (Equation 8). This allows $H$ to be expressed as a sum of these rank-one matrices ($E_k$).

Given that H was the transposed covariance of a part of the EEG, the maximum number of sums is equal to the number of channels (k). This decomposition method does not yield unique solutions. However, in ALCMV, to construct $E_i$, we kept the i-th column of matrix H and set the other elements to zero. The updated inverse covariance $C_{k+1}^{-1}$ is calculated by Equation 7.

$$H = E_1 + E_2 + \cdots + E_k$$

*Equation 11 step four in Miller equation*

$$C_{k+1} = G + E_1 + E_2 + \cdots + E_k$$

*Equation 12 step five in Miller equation*

$$v_k = \frac{1}{1 + tr(C_k^{-1} E_k)}$$

*Equation 13 step six in Miller equation*

*$v_k$: "tr" function sums the elements on the main diagonal of the square matrix, resulting in a one-dimensional $v_k$ (coefficient).*

$$(G + H)^{-1} = C_r^{-1} - v_r C_r^{-1} E_r C_r^{-1}$$

*Equation 14*

This method is capable of solving the inverse covariance recursively. We will have a recursive inverse covariance, which provides us with flexibility for quick responses. The covariance method, utilizing addition and subtraction with a function named "ICOV" in MATLAB, the general Miller method with a function titled "GMINVSUM", and the general miller inverse covariance with a function called

“RECUR_INVICOV”, have been implemented in MATLAB [13].

### D. Accelerated brain source orientation detection (AORI)

By examining the characteristics of the expression $L^T(r)R^{-1}L(r)$, specifically, we discovered that the result of $L^T(r)R^{-1}L(r)$ is always a 3×3 symmetric matrix. We used an optimal, reliable, fast, and dedicated method to calculate the eigenvector of this matrix, which is noteworthy for its minimal computations and ease of implementation. A custom function has been written for this purpose, available under the name “smlegnVec” [14]. The smlegnVec function has been implemented in MATLAB 2020a and achieves an accuracy equivalent to MATLAB commands with fewer computations. Fast orientation detection takes another step toward accelerated signal reconstruction of sources, as it reduces dimensionality by multiplying the lead field matrix, with dimensions (number of channels × 3), by the orientation matrix, with dimensions (3 × 1). This results in the involvement of a vector with dimensions (number of channels × 1) instead of a multidimensional matrix in future computations, significantly contributing to speed enhancement. In *Equation 9*, the maximum rank of the matrix $A=L^T(r)R^{-1}L(r)$ (number of independent rows or columns) will be 3. As a result, we will have a maximum of 3 eigenvalues (where i=1,2,3).

$$A = \begin{bmatrix} a_{11} & a_{12} & a_{13} \\ a_{21} & a_{22} & a_{23} \\ a_{31} & a_{32} & a_{33} \end{bmatrix}$$

*Equation 15 ‘A’ Matrix with elements*

In AORI, the rapid detection of brain source orientation was achieved through the use of Table 1, Table 2 (in appendix), and switch-case commands.

### E. Data acquisition

In addition to simulated EEG data, obtaining T1 MRI images and asynchronous 24-channel EEG from 4 healthy individuals aged 20 to 55 years. To generate EEG source signals, pre-prepared EEG data for voluntary movements of hands and feet available on the internet and publicly accessible have been used. Using Equation 16, we calculate the error for the number of brain sources in the x, y, and z directions relative to the maximum power method, and then normalize it by dividing by the maximum difference values. Where $[\,O_x\, O_y\, O_z\,]$ is the result of “eig” function and $[\hat{O}_x\, \hat{O}_y\, \hat{O}_z]$ as a result of “smlegnVec” function. We used absolute value cause both vector $[\hat{O}_x\, \hat{O}_y\, \hat{O}_z]$ and its opposite, $[\,-\hat{O}_x - \hat{O}_y - \hat{O}_z\,]$, are the vectors corresponding to the smallest eigenvalue. Therefore, both of them are correct solutions [15].

$$Ori_{error} = \frac{\left||\hat{O}_x| - |O_x|\right| + \left||\hat{O}_y| - |O_y|\right| + \left||\hat{O}_z| - |O_z|\right|}{3\max\left(\left||\hat{O}_x| - |O_x|\right|, \left||\hat{O}_y| - |O_y|\right|, \left||\hat{O}_z| - |O_z|\right|\right)}$$

*Equation 16*

With the presence of the direction vector, our signal has been reduced from 3 dimensions to 1 dimension. Therefore, for error analysis, we use Equation 17. In Equation 17, we calculate the absolute difference between the reconstructed signals of the accelerated and traditional LCMV methods and then normalized it with maximum value in each signal. N is the total number of elements in the reconstructed matrix, where r denotes the number of rows and c denotes the number of columns, as shown in Equation 18 [16] [17].

$$SignalRecons_{error} = \frac{\sum \left||\hat{S}_{ALCMV}| - |\hat{S}_{LCMV}|\right|}{N \left\||\hat{S}_{ALCMV}| - |\hat{S}_{LCMV}|\right\|_{max}}$$

*Equation 17*

$$N = r \times c$$

*Equation 18*

## III. RESULTS

The following sections present the results from applying the ALCMV, AORI, ORI, and LCMV methods to simulated EEG data with EEGg and real EEG with moving thumb movement task [18]. The orientation and signal reconstruction results are compared across the different approaches.

### A. ORI and LCMV Method Results

The traditional ORI method was applied to the same EEGg data to obtain orientation results, which were then compared with those from the AORI method. Additionally, the conventional LCMV (Linearly Constrained Minimum Variance) beamforming method was used for signal reconstruction. The results from LCMV were compared with the source reconstruction outputs from the ALCMV method. Errors for the orientation results were calculated using Equation 16, while errors for signal reconstruction were computed using Equation 16 and Equation 17.

### B. AORI, ALCMV and Comparative Results

EEGg (generated EEG) data were first processed using the AORI (Accelerated Brain Source Orientation Detection) method, which provided the initial orientation results. Following this, the ALCMV (Accelerated Linear Constrained Minimum Variance) method was applied to the EEGg data to perform source signal reconstruction. The reconstruction error was then calculated using Equation 16 and Equation 17. On average, ALCMV combined with AORI reduces computational load by 66% compared to LCMV and ORI. Additionally, the $Ori_{error}$ was consistently below 0.2%, indicating an accuracy of approximately 99.8%. The average $SignalRecons_{error}$ was 2%.

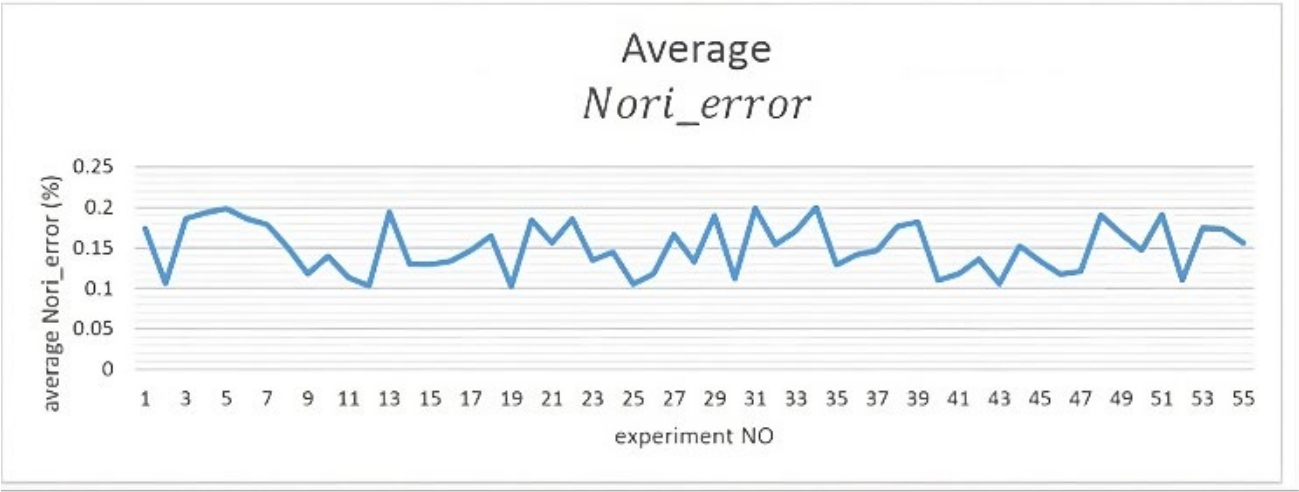

*Figure 2 AORI error compare to traditional ORI for EEGg signals*

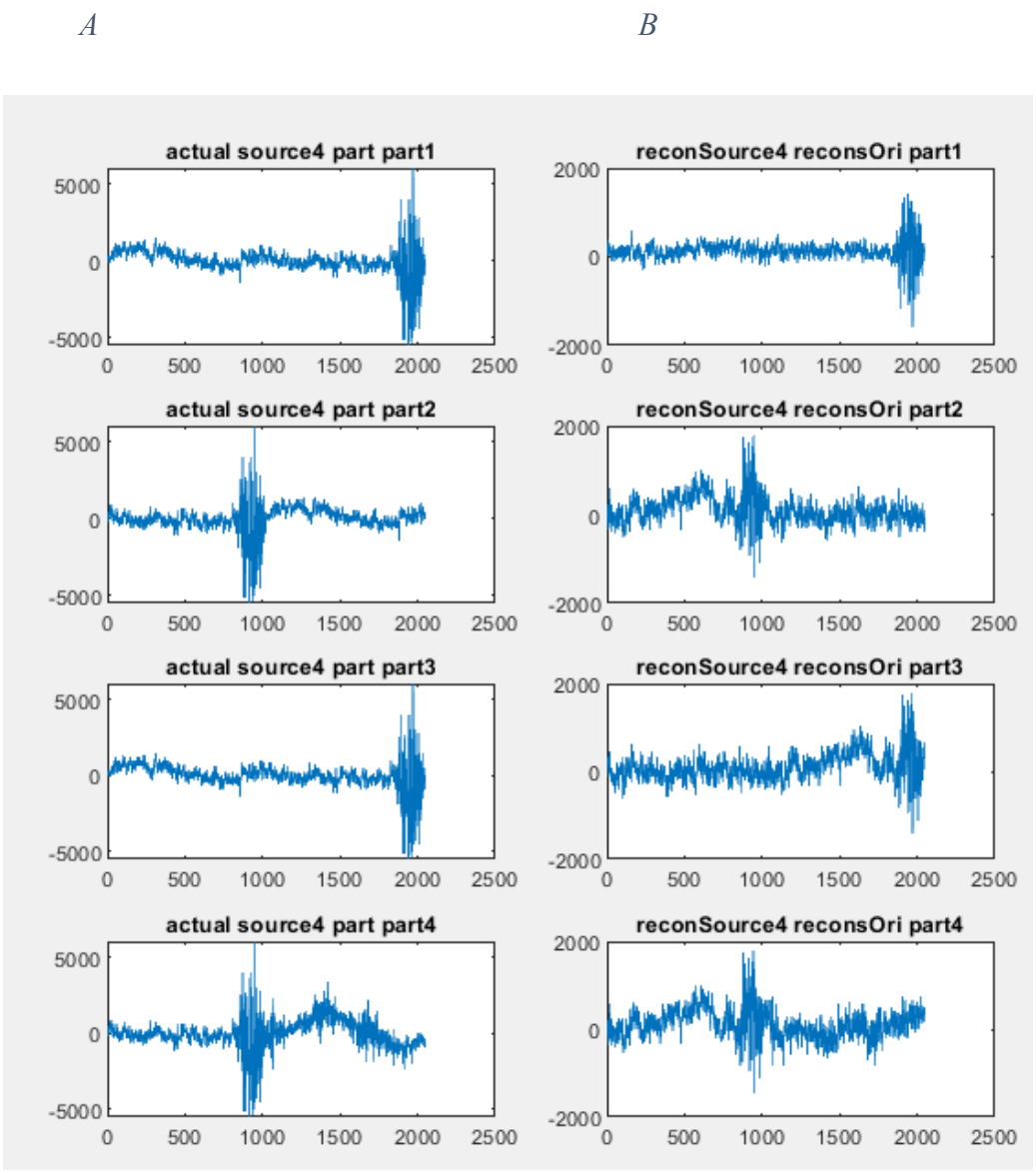

*Figure 3 Visualization of real sources and their reconstructions. 'A' column is the produced sources in EEGg. 'B' column is result of ALCMV and AORI.*

## C. Real EEG Validation

The most active source locations in the sensory motor cortex, identified by summing the absolute values of source signals, were detected using the ORI and LCMV methods, as well as their adaptive versions (AORI and ALCMV) [19] . These methods were validated using real EEG data. The detection error for the most active and second most active sources in 3D source localization was 3.2%, corresponding to an accuracy of 96.8%, with a 66% reduction in computational cost.

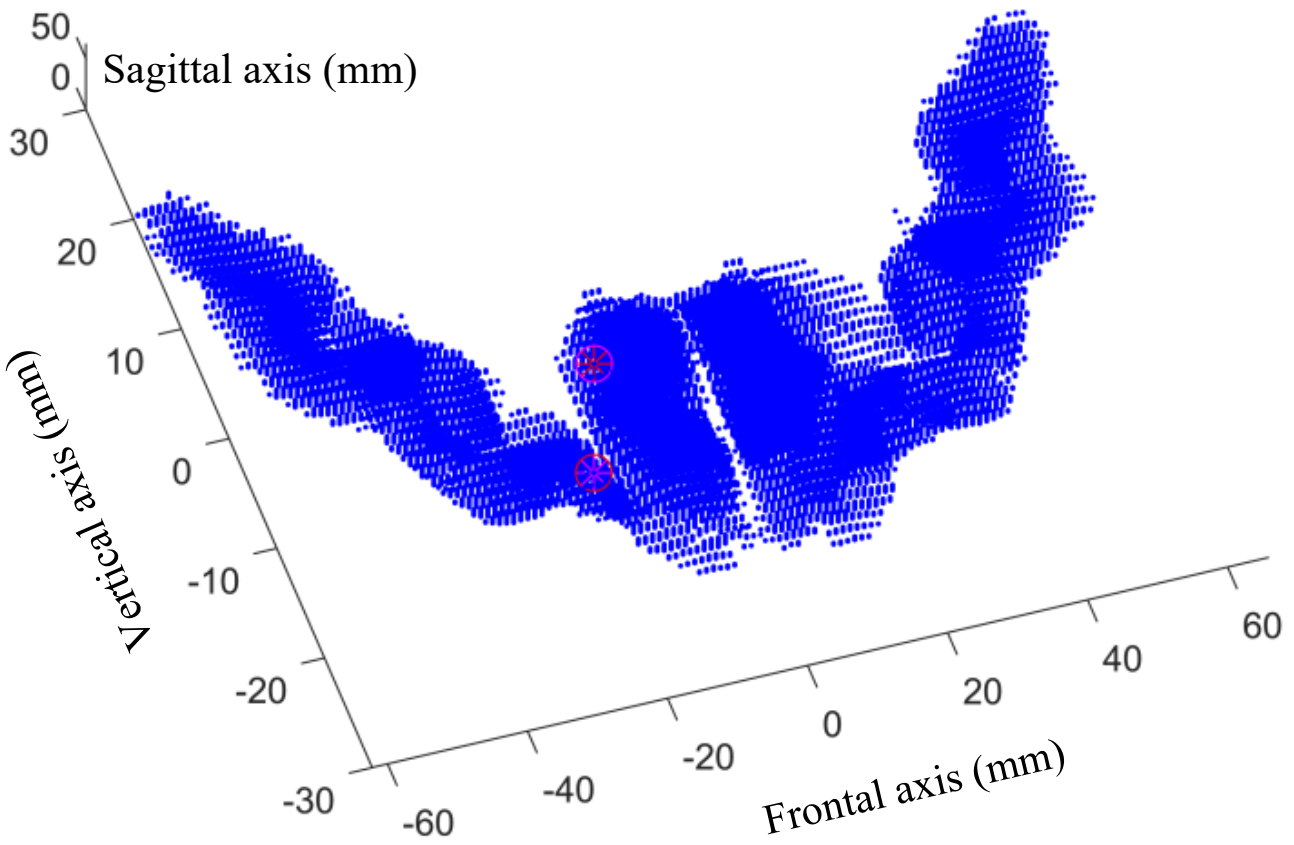

*Figure 4 Visualization of the most and the second most active sources*

In summary, the average reduction in computational load was 66%. The accuracy of detecting active sources from non-active sources was 99.92%, meaning that out of 10,000 sources, 9,992 were correctly classified as active or non-active. The average normalized correlation of signal reconstruction with EEG data was 0.98, while for the LCMV, ORI, ALCMV, and AORI methods, the average normalized correlation was 0.95. The average localization error for detecting the most active source using ALCMV and LCMV was 2.6 mm.

# IV. DISCUSSION

The results of this study highlight the significant performance improvements achieved through the proposed ALCMV and AORI algorithms in the context of EEG source localization. The primary advantage of these methods lies in their ability to reduce computational load while maintaining a high degree of accuracy. This is particularly important for real-time applications, such as neurorehabilitation and brain-computer interface (BCI) systems, where fast and accurate source localization is crucial.

The recursive covariance calculation in the ALCMV toolbox is a notable innovation, allowing the beamforming process to begin before the full set of data is available. This approach, combined with the General Miller method for inverse covariance calculation, significantly reduces the time required for source signal reconstruction. Similarly, the AORI toolbox simplifies the traditionally complex task of source orientation detection by reducing the problem from three dimensions to one. This not only accelerates the orientation estimation process but also maintains the accuracy of source localization.

The validation results from both simulated and real EEG data further reinforce the effectiveness of these methods. The 66% reduction in computational load is a substantial improvement over traditional techniques, such as LCMV and ORI, which often struggle with processing delays, especially when dealing with large datasets. Additionally, the low orientation error (below 0.2%) and signal reconstruction error (approximately 2%) demonstrate that these accelerated algorithms do not compromise the precision of EEG source localization.

In real-world applications, the enhanced speed and accuracy of the ALCMV and AORI toolboxes have significant implications. For example, in clinical diagnostics, fast and precise localization of abnormal brain activity could aid in early detection and intervention for neurological disorders, such as epilepsy. In BCI applications, the ability to localize brain signals in real time could improve the responsiveness and reliability of the system, thereby enhancing the overall user experience.

While this study focuses on EEG data, the principles behind the ALCMV and AORI algorithms could potentially be adapted for other types of neuroimaging data, such as magnetoencephalography (MEG). Future work could explore the application of these accelerated algorithms in other domains and further optimize the toolboxes for use in different EEG system configurations.

In conclusion, the ALCMV and AORI toolboxes represent a significant advancement in EEG source localization, providing a solution that balances speed and accuracy. These toolboxes offer practical benefits for both research and clinical settings, where real-time data processing is increasingly becoming a necessity.

### A. Limitations

The ALCMV system remains highly sensitive to the precise placement of EEG electrodes and their alignment with MRI images, making it dependent on laboratory conditions and requiring stricter standards for accurate results. It is also reliant on MRI data and the leadfield matrix [20], preventing it from functioning independently of these. Additionally, ALCMV is unable to reconstruct multiple source signals simultaneously, requiring significant computational resources for parallel processing. Its accuracy diminishes in the presence of background noise correlated with the dominant signal [21], and while higher spatial resolution is possible with more sources, it demands more powerful hardware and careful lab setups.

### B. Future Work

Future work could involve expanding non-invasive experiments to include more active sources, multiple tasks, and specialized cases such as brain diseases or tumors, enabling real-time tracking of brain activity. Simplifying the preparation of large datasets for deep learning through ALCMV and AORI could lead to real-time algorithmic diagnostic systems for patient classification. There is also potential for applying ALCMV in detecting readiness potential (RP) signals in intelligent machines [22] with BCIs, as well as integrating these algorithms for simultaneous fMRI and EEG recordings. Additionally, enhancing BCI systems and robotics with ALCMV and AORI for fast brain activity tracking could improve real-time control and responsiveness.

## V. Conclusion

In this study, we developed and evaluated two accelerated algorithms, ALCMV and AORI, aimed at enhancing the efficiency of EEG source localization. The results demonstrated that these algorithms significantly reduce computational load by 66% while maintaining high accuracy, with orientation and signal reconstruction errors below 0.2% and 2%, respectively. These improvements make the algorithms suitable for real-time applications, such as neurorehabilitation and brain-computer interfaces. Future work could focus on extending these methods to other neuroimaging techniques and optimizing them for broader clinical use.

## Acknowledgment

This study is part of an MSc thesis supported by the Tehran University of Medical sciences [code of ethics: IR.TUMS.MEDICINE.REC.1400.1102]. The codes are available: https://github.com/Avayekta/ALCMV-AORI.git

APPENDIX

*Table 1 Optimal Computation of Eigenvectors: Part One [14]*

| eign vector of symmetric matrix | | |
|---|---|---|
| **case** | **eign vectors** | |
| Only Diagonal Elements Present in A | {{1,0,0},{0,1,0},{0,0,1}} | |
| If $\mathbf{A}=\begin{bmatrix} a_{11} & 0 & 0 \\ 0 & a_{22} & a_{23} \\ 0 & a_{23} & a_{33} \end{bmatrix}$ 2.1 | $\left\{\{1,0,0\},\left\{0,-\frac{b_{23}}{\sqrt{b_{22}^{i2}+b_{23}^2}},\frac{b_{22}^i}{\sqrt{b_{22}^{i2}+b_{23}^2}}\right\},\left\{0,-\frac{b_{22}^i}{\sqrt{b_{22}^{i2}+b_{23}^2}},-\frac{b_{23}}{\sqrt{b_{22}^{i2}+b_{23}^2}}\right\}\right\}$ | |
| If $\mathbf{A}=\begin{bmatrix} a_{11} & 0 & a_{13} \\ 0 & a_{22} & 0 \\ a_{13} & 0 & a_{33} \end{bmatrix}$ 2.2 | $\left\{\left\{-\frac{b_{13}}{\sqrt{b_{11}^{i2}+b_{13}^2}},0,\frac{b_{11}^i}{\sqrt{b_{11}^{i2}+b_{13}^2}}\right\},\{0,1,0\},\left\{\frac{b_{11}^i}{\sqrt{b_{11}^{i2}+b_{13}^2}},0,\frac{b_{13}}{\sqrt{b_{11}^{i2}+b_{13}^2}}\right\}\right\}$ | |
| If $\mathbf{A}=\begin{bmatrix} a_{11} & a_{12} & 0 \\ a_{12} & a_{22} & 0 \\ 0 & 0 & a_{33} \end{bmatrix}$ 2.3 | $\left\{\left\{-\frac{b_{12}}{\sqrt{b_{11}^{i2}+b_{12}^2}},\frac{b_{11}^i}{\sqrt{b_{11}^{i2}+b_{12}^2}},0\right\},\left\{-\frac{b_{11}^i}{\sqrt{b_{11}^{i2}+b_{12}^2}},-\frac{b_{12}}{\sqrt{b_{11}^{i2}+b_{12}^2}},0\right\},\{0,0,1\}\right\}$ | |
| *If* $(b_{11}^i b_{23}-b_{13}b_{12})b_{13}\neq 0$ *or* $(b_{12}^2-b_{11}^i b_{22}^i)b_{13}\neq 0$ 3.1 | $\left\{\frac{\mathcal{P}_i^n}{\sqrt{\mathcal{P}_i^{n2}+\mathcal{Q}_i^2+1}},\frac{\mathcal{Q}_i}{\sqrt{\mathcal{P}_i^{n2}+\mathcal{Q}_i^2+1}},\frac{1}{\sqrt{\mathcal{P}_i^{n2}+\mathcal{Q}_i^2+1}}\right\}$ | $\mathcal{Q}_i=\frac{b_{11}^i b_{23}-b_{13}b_{12}}{b_{12}^2-b_{11}^i b_{22}^i}$ $\mathcal{R}_i=\frac{b_{12}^2-b_{11}^i b_{22}^i}{b_{11}^i b_{23}-b_{13}b_{12}}$ $\mathcal{P}_i^n=-\frac{b_{23}\mathcal{Q}_i+b_{33}^i}{b_{13}}$ |
| If $(b_{11}^i b_{33}^i-b_{13}^2)b_{12}\neq 0$ **or** $(b_{12}b_{13}-b_{11}^i b_{23})b_{12}\neq 0$ 3.2 | $\left\{\frac{\mathcal{P}_i^n}{\sqrt{\mathcal{P}_i^{n2}+\mathcal{Q}_i^2+1}},\frac{\mathcal{Q}_i}{\sqrt{\mathcal{P}_i^{n2}+\mathcal{Q}_i^2+1}},\frac{1}{\sqrt{\mathcal{P}_i^{n2}+\mathcal{Q}_i^2+1}}\right\}$ | $\mathcal{Q}_i=\frac{b_{11}^i b_{33}^i-b_{13}^2}{b_{12}b_{13}-b_{11}^i b_{23}}$ $\mathcal{R}_i=\frac{b_{12}b_{13}-b_{11}^i b_{23}}{b_{11}^i b_{33}^i-b_{13}^2}$ $\mathcal{P}_i^n=-\frac{b_{22}^i\mathcal{Q}_i+b_{23}}{b_{12}}$ |
| If $(b_{12}b_{33}^i-b_{23}b_{13})b_{11}^i\neq 0$ **or** $(b_{22}^i b_{13}-b_{12}b_{23})b_{11}^i\neq 0$ 3.3 | $\left\{\frac{\mathcal{P}_i^n}{\sqrt{\mathcal{P}_i^{n2}+\mathcal{Q}_i^2+1}},\frac{\mathcal{Q}_i}{\sqrt{\mathcal{P}_i^{n2}+\mathcal{Q}_i^2+1}},\frac{1}{\sqrt{\mathcal{P}_i^{n2}+\mathcal{Q}_i^2+1}}\right\}$ | $\mathcal{Q}_i=\frac{b_{12}b_{33}^i-b_{23}b_{13}}{b_{22}^i b_{13}-b_{12}b_{23}}$ $\mathcal{R}_i=\frac{b_{22}^i b_{13}-b_{12}b_{23}}{b_{12}b_{33}^i-b_{23}b_{13}}$ $\mathcal{P}_i^n=-\frac{b_{12}\mathcal{Q}_i+b_{13}}{b_{11}^i}$ |

*Table 2 Optimal Computation of Eigenvectors: Part Two [14]*

| eign vector of symmetric matrix | |
|---|---|
| **case** | **eign vectors** |
| If $(b_{12}b_{23} - b_{13}b_{22}^i)b_{23} \neq 0$<br>**or** $(b_{11}^i b_{22}^i - b_{12}^2)b_{23} \neq 0$<br><br>**3.4** | $\left\{\frac{\mathcal{P}_i}{\sqrt{\mathcal{P}_i^2+\mathcal{Q}_i^{n2}+1}}, \frac{\mathcal{Q}_i^n}{\sqrt{\mathcal{P}_i^2+\mathcal{Q}_i^{n2}+1}}, \frac{1}{\sqrt{\mathcal{P}_i^2+\mathcal{Q}_i^{n2}+1}}\right\}$<br>$\mathcal{P}_i = \frac{b_{12}b_{23} - b_{13}b_{22}^i}{b_{11}^i b_{22}^i - b_{12}^2}$<br>$\mathcal{R}_i = \frac{b_{11}^i b_{22}^i - b_{12}^2}{b_{12}b_{23} - b_{13}b_{22}^i}$<br>$\mathcal{Q}_i^n = -\frac{b_{13}\mathcal{P}_i + b_{33}^i}{b_{23}}$ |
| If $(b_{12}b_{33}^i - b_{13}b_{23})b_{22}^i \neq 0$<br>or $(b_{11}^i b_{23} - b_{12}b_{13})b_{22}^i \neq 0$<br><br>**3.5** | $\left\{\frac{\mathcal{P}_i}{\sqrt{\mathcal{P}_i^2+\mathcal{Q}_i^{n2}+1}}, \frac{\mathcal{Q}_i^n}{\sqrt{\mathcal{P}_i^2+\mathcal{Q}_i^{n2}+1}}, \frac{1}{\sqrt{\mathcal{P}_i^2+\mathcal{Q}_i^{n2}+1}}\right\}$<br>$\mathcal{P}_i = \frac{b_{12}b_{33}^i - b_{13}b_{23}}{b_{11}^i b_{23} - b_{12}b_{13}}$<br>$\mathcal{R}_i = \frac{b_{11}^i b_{23} - b_{12}b_{13}}{b_{12}b_{33}^i - b_{13}b_{23}}$<br>$\mathcal{Q}_i^n = -\frac{b_{12}\mathcal{P}_i + b_{23}}{b_{22}^i}$ |
| If $(b_{22}^i b_{33}^i - b_{23}^2)b_{12} \neq 0$<br>or $(b_{12}b_{23} - b_{22}^i b_{13})b_{12} \neq 0$<br><br>**3.6** | $\left\{\frac{\mathcal{P}_i}{\sqrt{\mathcal{P}_i^2+\mathcal{Q}_i^{n2}+1}}, \frac{\mathcal{Q}_i^n}{\sqrt{\mathcal{P}_i^2+\mathcal{Q}_i^{n2}+1}}, \frac{1}{\sqrt{\mathcal{P}_i^2+\mathcal{Q}_i^{n2}+1}}\right\}$<br>$\mathcal{P}_i = \frac{b_{22}^i b_{33}^i - b_{23}^2}{b_{12}b_{23} - b_{22}^i b_{13}}$<br>$\mathcal{R}_i = \frac{b_{12}b_{23} - b_{22}^i b_{13}}{b_{22}^i b_{33}^i - b_{23}^2}$<br>$\mathcal{Q}_i^n = -\frac{b_{11}^i\mathcal{P}_i + b_{13}}{b_{12}}$ |
| If $(b_{13}b_{22}^i - b_{12}b_{23})b_{33}^i \neq 0$<br>or $(b_{11}^i b_{23} - b_{13}b_{12})b_{33}^i \neq 0$<br><br>**3.7** | $\left\{\frac{\mathcal{P}_i}{\sqrt{\mathcal{P}_i^2+1+\mathcal{R}_i^{m2}}}, \frac{1}{\sqrt{\mathcal{P}_i^2+1+\mathcal{R}_i^{m2}}}, \frac{\mathcal{R}_i^m}{\sqrt{\mathcal{P}_i^2+1+\mathcal{R}_i^{m2}}}\right\}$<br>$\mathcal{P}_i = \frac{b_{13}b_{22}^i - b_{12}b_{23}}{b_{11}^i b_{23} - b_{13}b_{12}}$<br>$\mathcal{Q}_i = \frac{b_{11}^i b_{23} - b_{13}b_{12}}{b_{13}b_{22}^i - b_{12}b_{23}}$<br>$\mathcal{R}_i^m = -\frac{b_{13}\mathcal{P}_i + b_{23}}{b_{33}^i}$ |
| If $(b_{13}b_{23} - b_{12}b_{33}^i)b_{23} \neq 0$<br>or $(b_{11}^i b_{33}^i - b_{13}^2)b_{23} \neq 0$<br><br>**3.8** | $\left\{\frac{\mathcal{P}_i}{\sqrt{\mathcal{P}_i^2+1+\mathcal{R}_i^{m2}}}, \frac{1}{\sqrt{\mathcal{P}_i^2+1+\mathcal{R}_i^{m2}}}, \frac{\mathcal{R}_i^m}{\sqrt{\mathcal{P}_i^2+1+\mathcal{R}_i^{m2}}}\right\}$<br>$\mathcal{P}_i = \frac{b_{13}b_{23} - b_{12}b_{33}^i}{b_{11}^i b_{33}^i - b_{13}^2}$<br>$\mathcal{Q}_i = \frac{b_{11}^i b_{33}^i - b_{13}^2}{b_{13}b_{23} - b_{12}b_{33}^i}$<br>$\mathcal{R}_i^m = -\frac{b_{12}\mathcal{P}_i + b_{22}^i}{b_{23}}$ |
| If $(b_{23}^2 - b_{22}^i b_{33}^i)b_{13} \neq 0$<br>or $(b_{12}b_{33}^i - b_{23}b_{13})b_{13} \neq 0$<br><br>**3.9** | $\left\{\frac{\mathcal{P}_i}{\sqrt{\mathcal{P}_i^2+1+\mathcal{R}_i^{m2}}}, \frac{1}{\sqrt{\mathcal{P}_i^2+1+\mathcal{R}_i^{m2}}}, \frac{\mathcal{R}_i^m}{\sqrt{\mathcal{P}_i^2+1+\mathcal{R}_i^{m2}}}\right\}$<br>$\mathcal{P}_i = \frac{b_{23}^2 - b_{22}^i b_{33}^i}{b_{12}b_{33}^i - b_{23}b_{13}}$<br>$\mathcal{Q}_i = \frac{b_{12}b_{33}^i - b_{23}b_{13}}{b_{23}^2 - b_{22}^i b_{33}^i}$<br>$\mathcal{R}_i^m = -\frac{b_{11}^i\mathcal{P}_i + b_{12}}{b_{13}}$ |